\documentstyle[epsf,psfig,prb,twocolumn,aps]{revtex}

\begin{document}

\draft
\input epsf
\twocolumn[\hsize\textwidth\columnwidth\hsize\csname
@twocolumnfalse\endcsname

\title{A critical comparison of classical 
field theory and microscopic wavefunctions 
for skyrmions in quantum Hall ferromagnets}
\author{M. Abolfath$^{1}$,J. J. Palacios$^{1,2}$, H. A. Fertig$^{2}$, S. M.
Girvin$^{1}$, and A. H. MacDonald$^{1}$}
\address{$^{1}$ Department of Physics, Indiana University, Bloomington, IN
47405}
\address{$^{2}$ Department of Physics and Astronomy, University of Kentucky,
Lexington, KY 40506} 

\date{\today}

\maketitle

\widetext
\begin{abstract}
\leftskip 2cm
\rightskip 2cm

We report on a study of the classical field theory description of
charged skyrmions in quantum Hall ferromagnets. The appropriate field
theory is a non-linear $\sigma$ model generalized to include Coulomb and
Zeeman interaction terms. We have tested the range of validity of the
classical field theory by comparing with microscopic descriptions of the
single skyrmion state based on the Hartree-Fock approximation, exact
diagonalization calculations, and many-body trial wavefunctions. We find
that the field theory description is accurate for skyrmions with
moderate spin quantum numbers ($\gtrsim 10$) although, as expected, it
fails qualitatively for small spin quantum numbers.  


\end{abstract}

\vskip2pc]

\narrowtext

\section{Introduction}

The quantum Hall effect\cite{smgbook,ahmbook,chakrabortybook,stonebook}
occurs in a two-dimensional electron system (2DES) in a strong
perpendicular magnetic field ($B$) and is associated with the existence of
incompressible ground states at certain values of the Landau level
filling factor, $\nu$. ($\nu \equiv N/N_{\phi}$ where $N_{\phi}$ is the
Landau level degeneracy and $N$ is the number of electrons in the
system.) Some incompressible ground states are strong ferromagnets,
i.e., they are total spin eigenstates with $S = N/2$ so that an
infinitesimal Zeeman coupling of the magnetic field to the spin degree
of freedom is sufficient to produce complete spin alignment. Because of
the small ratio of the Zeeman-splitting to other relevant energy scales
in typical 2DES's, it is then useful to regard the system as a
ferromagnet in a weak symmetry breaking magnetic field, even though the
2DES is often in the strong field quantum limit as far as orbital
degrees of freedom are concerned. Quantum Hall ferromagnets (QHF's) have
a number of unusual properties that spring from the disparity between
the magnetic field coupling strengths for orbital and spin degrees of
freedom.

Because of the gap for charged excitations in incompressible states, the
only low-lying excitations in these quantum Hall ferromagnets are those
associated with slow variations in the unit vector field which describes
the local orientation of the spin magnetic moment. In addition to the
spin waves modes, there exist higher energy topologically non-trivial
skyrmion\cite{kaneandlee,Sondhi} textures in the spin field. The
importance of these topologically non-trivial excitations is magnified
by the fact that they carry an electrical charge.\cite{Sondhi,Moon} As
the lowest energy charge carriers they control the thermally activated
dissipation on the $\nu=1$ quantum Hall plateau and, away from filling
factor $\nu = 1$, they are present in the ground
state.\cite{Herb,hcm,Brey} In the latter case the ground state is no
longer fully spin polarized.  For weak Zeeman coupling skyrmions are
large and each charge introduced into the system flips over a large
number of spins. Thus the spin magnetization is expected to show a sharp
cusp at filling factor $\nu=1$.

The presence of these objects has recently been detected directly in NMR
measurements of the electron spin magnetization which is proportional to
the Knight shift\cite{barrett}, in thermal transport
measurements\cite{jpe-skyrme}, and in optical absorption.\cite{optical1}
They may also be responsible for the recently observed\cite{bayot}
enormous enhancement of the apparent specific heat because of their
equilibrating effect on the nuclear spins. There has recently been
considerable interest in the physics of these exotic electronic
quasiparticles.
\cite{kaneandlee,Sondhi,Moon,Herb,hcm,Brey,kamilla,jacob,karlhede,%
damping,statistics,wilczek,stone,jasonho,tsvelik,wu,jjp:xminus}

Charged spin texture excitations in QHF's are the subject of this paper.
A realistic theory of these excitations requires at a minimum that
Coulomb energy and Zeeman energy terms be added to the non-linear sigma
(NL$\sigma$) model classical field theory. In Sec.\
\ref{sec:fieldtheory} we briefly review the resulting classical field
theory of spin textures in QHF's and report on numerical solutions for
the lowest energy topologically non-trivial textures.  With typical
parameters, the size of the lowest energy textures need not be large
compared to microscopic lengths so that both higher order gradient
corrections and quantum fluctuations not included in the model may
become important. In Sec.\ \ref{sec:micro} we present an overview of the
various microscopic approaches which have been used to study skyrmion
states in quantum Hall ferromagnets. In the Hartree-Fock approximation,
the gradient expansion of the field theory is effectively summed to
infinite order. However, it becomes difficult to solve these equations
accurately for very large skyrmions.  Many-particle variational
wavefunction and exact diagonalization approaches also incorporate, in
addition, quantum fluctuation effects, but are even more limited in the
size of quasiparticles for which accurate calculations are possible.
Numerical comparisons of energies and spin densities obtained using
these different approximation strategies are compared in Sec.\
\ref{sec:comparison}.  One conclusion from this work is that the most
important corrections to the minimal generalized NL$\sigma$ model are
the higher order gradient corrections already captured in the
Hartree-Fock approximation.  The quantum fluctuations added using trial
wavefunction or exact diagonalization are responsible only for minor
further modifications. We summarize our conclusions in Section
\ref{sec:concl}.

\section{Classical field theory}
\label{sec:fieldtheory}

\subsection{General considerations}

Here we briefly review the classical field theory for charged spin texture
excitations or skyrmions in quantum Hall ferromagnets.\cite{Sondhi,Moon}
Following Sondhi {\em et al.},\cite{Sondhi} we start from a modified version of
the NL$\sigma$ model classical
 field theory which has been exploited to describe
the low energy properties of two-dimensional Heisenberg ferromagnets and
antiferromagnets.\cite{Polyakov,Haldane,Auerbach1,Auerbach} The order parameter
in this theory is a unit vector field, ${\bf m}({\bf r})$, that describes the
local orientation of the spin or pseudospin\cite{Moon} magnetic order. For
quantum Hall ferromagnets the energy functional of the minimal theory is 
\begin{equation}
E[{\bf m}]=E_0[{\bf m}]+E_z[{\bf m}]+E_c[{\bf m}] \; ,
\label{eq1}
\end{equation}
where $E_0[{\bf m}]$ is the leading order term in a gradient expansion of the
energy functional and $E_z[{\bf m}]$ is the Zeeman energy functional:
\begin{mathletters}
\label{eq2}
\begin{equation}
E_0[{\bf m}]=\frac {\rho_s}{2}\; \int d^{2}r\, (\nabla{\bf m})^2\; , 
\label{eq2a}
\end{equation}
\begin{equation}
E_z[{\bf m}]=\frac{t}{2 \pi \ell_0^2} \int d^{2}r\, [ \; 1- m_z({\bf r}) \; ].
\label{eq2b}
\end{equation}
Here $\rho_s=e^2/(16\sqrt{2\pi}\epsilon\ell_0)$ is the spin stiffness (assuming
zero layer thickness for the 2DEG)
 which can be calculated analytically for this
case and represents a loss of Coulomb exchange energy, $t= (g^* \mu_B B)/2$
represents the Zeeman coupling strength, $\epsilon$ is the dielectric constant
of the host semiconductor, and $\ell_0$ is the magnetic length.

When only these first two terms are present, the spatial extent of structure in
skyrmion extrema of the energy functional shrinks to a point and the gradient
expansion fails. To obtain a consistent theory of skyrmion excitations when the
Zeeman coupling is non-zero it is necessary to go beyond leading order in the
gradient expansion. For QHF's it turns out\cite{Sondhi} that, because skyrmions
are charged, the next to leading term in
 the gradient expansion is the non-local
Coulomb interaction energy functional
\begin{equation}
E_c[{\bf m}]={e^2\over 2 \epsilon} \int d^{2}r \int d^{2}r^\prime \,
\frac{\rho({\bf r})\rho({\bf r}^\prime)}{|{\bf r}-{\bf r}^\prime|} \; .
\label{eq2c}
\end{equation}
\end{mathletters}
where the charge density is given by\cite{Sondhi,Moon}
\begin{equation}
\rho({\bf r})=\frac{-\nu}{8 \pi } \; \epsilon_{a b} \; {\bf m({\bf r})} \cdot
[\partial_a{\bf m({\bf r})} \wedge \partial_b{\bf m({\bf r})}] \; . 
\end{equation}
The charge density is the product of the Landau level filling factor of the
ground state $\nu$ and the $O(3)$ topological density of the spin texture. The
non-locality of this functional complicates the calculations we present below.
The total skyrmion charge is therefore
 $\nu$ times an integer-valued topological
invariant. If we compactify the 2D plane to a sphere $S^2$, then ${\bf m}({\bf
r})$ defines a mapping of $S^2$ onto the order parameter sphere $S^2$ and hence
can be classified by the homotopy group $\pi_2(S^2)$. The topological charge
simply counts the wrapping of the order parameter sphere by the coordinate
sphere.\cite{fradkin,rajaraman}

Extrema of the energy functional of this model, ${\bf \tilde{m}}$, satisfy a
non-linear differential equation which can be obtained by minimizing Eq.\
(\ref{eq1}) with respect to ${\bf m}$, using a Lagrange multiplier to enforce
the constraint $\bf m(\bf r) \cdot \bf m(\bf r) = 1$. After eliminating the
Lagrange multiplier we find that 
\begin{eqnarray}
\rho_s \Bigl( -\nabla^2 + {\bf \tilde{m}} \cdot \nabla^2 {\bf \tilde{m}}
\Bigl){\rm \tilde{m}}_\mu -\frac{t}{2\pi \ell_0} (\delta_{z\mu}- {\rm
\tilde{m}}_z{\rm \tilde{m}}_\mu) - & \nonumber \\
\frac{\nu}{4\pi} \epsilon_{ab} \{ \partial_a
V({\bf r}) \} ({\bf \tilde{m}} \wedge \partial_b {\bf \tilde{m}})_\mu&=0& \; ,
\label{eq3}
\end{eqnarray}
where $V({\bf r})$ is Hartree potential 
\begin{equation}
V({\bf r})=\frac{e^2}{\epsilon_0} \int d{\bf r}^\prime \frac{\tilde{\rho}({\bf
r}^\prime)} {|{\bf r}-{\bf r}^\prime|} \; ,
\label{eq4}
\end{equation}
and $\tilde{\rho}$ is the skyrmion charge density corresponding to the minimum
energy solution, ${\bf \tilde{m}}({\bf r})$. The solutions of Eq.\ (\ref{eq3})
can be classified by the skyrmion charge $Q=\int d{\bf r} \; \rho({\bf r})$. In
the absence of Zeeman and Coulomb energies the energy functional is scale
invariant and Eq.\ (\ref{eq3}) has a family of known analytic scale invariant
solutions.\cite{rajaraman,Blavin,Woo} The solutions may be represented by
analytic complex valued polynomials: 
\begin{equation}
{\bf \tilde{m}}(z) = \left(\frac{4w_1}{|w|^2+4}, \frac{4w_2}{|w|^2+4},
\frac{|w|^2-4}{|w|^2+4}\right) \; ,
\label{eq6_1}
\end{equation}
where $w = w_1 + i w_2$ is a polynomial in the complex variable
$z=(x+iy)/\lambda$ with arbitrary scale $\lambda$. The degree of the polynomial
$w$ determines the skyrmion charge. These analytic solutions are not valid if
the Zeeman and Coulomb terms are included. In the following subsection we
briefly explain the numerical methods we use to solve the generalized
differential equation for the case of unit charge spin texture.

\subsection{Single skyrmion}

In this subsection we concentrate our attention on antiskyrmions (skyrmions)
with unit topological charge, $Q=1 (-1)$. We find a numerical solution to Eq.\
(\ref{eq3}) for an infinite system with the boundary condition that the spin is
down at the skyrmion center and up at infinity. It is convenient to choose the
origin of coordinates at the center of the skyrmion to take advantage of the
circular symmetry of the charge distribution of a single skyrmion. We
parameterize the order
 parameter ${\bf m}({\bf r})$ by spherical angles so that 
\begin{equation}
{\bf m}({\bf r})= \left(\sin \theta({\bf r}) \cos \phi({\bf r}), 
                        \sin \theta({\bf r}) \sin \phi({\bf r}),
                        \cos \theta({\bf r}) \right) \; ,
\label{eq5}
\end{equation}
where $\theta({\bf r})$ and $\phi({\bf r})$ are new field variables. We take
advantage of the circular symmetry for the single-skyrmion problem by choosing
$\theta(r, \varphi) \equiv \theta(r)$ and $\phi(r, \varphi) \equiv \pm\varphi +
\varphi_0$ where $\varphi$ is the azimuthal 
angle and $\varphi_0$ is a constant.
The energy of the skyrmion is independent of $\varphi_0$ because of the
invariance of the energy functional under global rotation of all spins in the
plane about the $z$-axis. Positive and negative signs for $\varphi$ yield
antiskyrmion and skyrmion solutions, respectively. One may show explicitly that
the above ansatz and boundary conditions yield unit winding number, $Q=\pm1$,
since the charge density in the circularly symmetric case is given by
$\rho(r)=\mp(1/4\pi r)(d/dr \cos\theta(r)) $. It follows that skyrmion and
antiskyrmion solutions have the same energy, allowing us to consider skyrmion
spin textures only.\cite{phscaveat} In this case, Eq.\ (\ref{eq3}) reduces to a
non-linear integro-differential equation for $\theta (r)$: 
\begin{eqnarray}
&&\rho_s\frac{1}{r}\frac{d}{dr} \Bigl( r\frac{d\theta(r)}{dr} \Bigl)
-\rho_s\frac{1}{2r^2}\sin(2\theta(r)) -
\frac{t}{2\pi\ell_0^2}\sin\theta(r)\nonumber\\
&&+\frac{e^2}{16\pi^2}\frac{\sin\theta(r)}{r}
\int dr^\prime \frac{dU(r,r^\prime)}{dr}\sin\theta(r^\prime)
\frac{d\theta(r^\prime)}{dr^\prime}= 0 \; .
\label{eq6}
\end{eqnarray}
Here
\begin{equation}
U(r,r^\prime)\equiv\frac{4}{|r-r^\prime|} K[\frac{(-4 r
r^\prime)}{(r-r^\prime)^2}]
\end{equation} 
where $K(x)$ is the complete Elliptic integral of the first kind
continued appropriately to negative values of $x$. We have solved Eq.\
(\ref{eq6}) for $\theta (r)$ using an iterative approach. At each step
in the calculation the integral over $r'$ was evaluated at each $r$
using an approximation for $\theta (r)$. When this integral is fixed, we
are left with a two-point boundary value problem [$ \theta(r=0) = \pi$
and $\theta(r \to \infty) = 0$] which can be solved by standard methods.
The resulting value of $\theta (r)$ to evaluate the integral over $r'$
for the next iteration. We found that this iterative procedure converged
rapidly when the iteration was started from one of the analytic
solutions obtained neglecting Coulomb and Zeeman energies.

\begin{figure}
\centerline{\epsfxsize=8cm \epsfbox{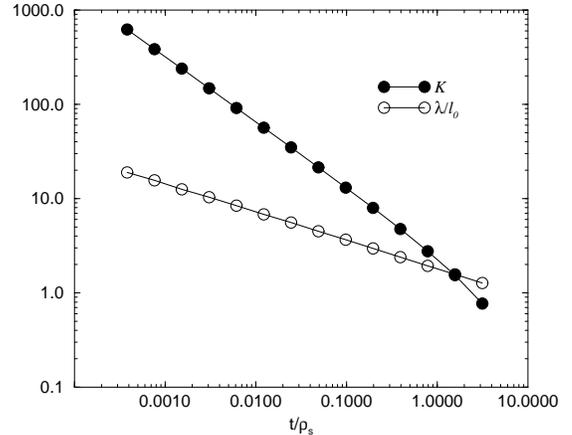}}
\caption{Skyrmion size (circles) and spin (black dots) obtained from the
classical field theory as a function of the Zeeman coupling constant $t$. The
skyrmion size $\lambda$ is defined as the radius at
which the spin lies in the xy plane,
i.e., $\cos[\theta(\lambda)]=0$.  These results are consistent with
the power law $\propto t^{-1/3}$ behavior expect at small $t$. The number
of reversed spins in a Skyrmion is proportional to $\lambda^2 \propto
t^{-2/3}$.  Its deviation from these power law at large $t$
is an indication of the necessity for including microscopic physics
not captured by the minimal field theory.}
\label{ftfig1}
\end{figure}

Figure \ref{ftfig1} illustrates some of the results obtained from these
numerical calculations. We plot the size of the skyrmion ($\lambda$), 
defined by
$\theta (\lambda) = \pi/2$, and the the number of reversed spins ($K$), defined
by 
\begin{equation}
K \equiv \frac{1}{4\pi\ell_0^2} \int d {\bf r} [ 1 - m_z({\bf r}) ] -
\frac{1}{2},
\label{eqK}
\end {equation}
as a function of $t$. We find that $ \lambda \sim t^{-1/3}$ and $K \sim
t^{-2/3}$. The logarithmic corrections to these power laws predicted on the
basis of a variational solution of the field theory\cite{Sondhi} are not yet
apparent at the smallest values of $t$ we have considered.

\begin{figure}
\centerline{\epsfxsize=8cm \epsfbox{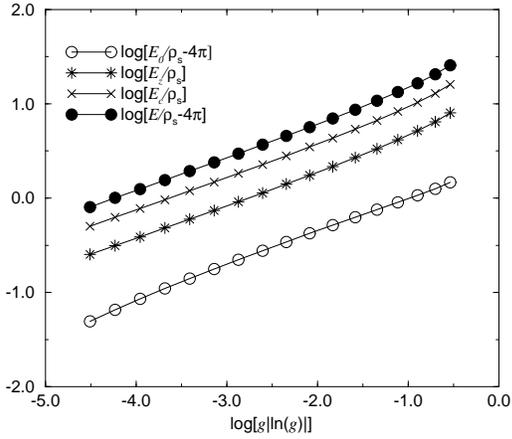}}
\caption{Skyrmion
energies: $E_c$ (crosses), $E_z$ (stars), excess gradient energy
$E_0-4\pi$ (circles), and total excess energy (black dots) as a function
of log[$g$ln($g$)] [g is the Zeeman spin splitting in units of $e^2/(\epsilon
\ell_0$)]. The power law for the dependence of
the excess skyrmion energy, which we extract from a
linear fit to these plots, is close to 1/3.}
\label{ftfig2}
\end{figure}

Figure \ref{ftfig2} presents our numerical results for the dependence of
skyrmion energy on Zeeman coupling strength in the minimal field
theoretical description. The Blavin-Polyakov solutions of Eq.\
(\ref{eq6_1}) have the minimum possible gradient energy, which has the
value $4\pi\rho_s$. When the Zeeman and the Coulomb terms are added the
skyrmion texture is reshaped and the optimal solutions are no longer
analytic in $z$. In particular\cite{Sondhi} the component of the
magnetization perpendicular to the Zeeman field has an exponential
rather than a logarithmic fall of at large distances. These non-analytic
solutions must have a gradient energy larger than $4\pi\rho_s$, with the
deviation from this value decreasing as the Zeeman energy decreases, or,
equivalently, as the size of the skyrmion increases. We thus expect that
for the gradient energy $E_0(\lambda) = 4\pi\rho_s + f(\lambda)$ where
$f(\lambda)$ is monotonically decreasing function of the skyrmion's size
(see Fig.\ \ref{ftfig2}). The non-analytical solutions become
asymptotically close to the Blavin-Polyakov solutions at small Zeeman
splitting (large $\lambda$).  Dimensional analysis shows that, up to
possible logarithmic factors, the Coulomb and the Zeeman energies must
vary at large $\lambda$ as $C/\lambda$ and $D \lambda^2$, respectively,
where $C \propto e^2/\epsilon $ and $D \propto t/\ell_0^2$ 
are constants. Assuming that
$f(\lambda)$ vanishes faster than $1/\lambda$, optimizing the total
energy $[4\pi\rho_s+f(\lambda)+C/\lambda+D\lambda^2]$ with respect to
$\lambda$ leads to the prediction $E_z = E_c/2 \propto (e^2/\epsilon
\ell_0)^{2/3} t^{1/3}$ as
$\lambda \rightarrow \infty$. These power laws were originally
suggested\cite{Sondhi} by Sondhi {\em et al.} Our numerical results
confirm these predictions, which hold with surprising accuracy out to
fairly large values of $t$: in fact $E_z/E_c$ is within 1\% of its
asymptotic value in the whole range of values of $\lambda$ we have
studied. The fact that $f(\lambda)$ vanishes faster than $\lambda^{-1}$
is expected on the basis of perturbative treatments of the Coulomb and
Zeeman terms and confirmed by the numerical results shown in Figure
\ref{ftfig2}. The total energy of the Skyrmion is closely given by $
E/\rho_s = 4 \pi + A [g |\ln (g) |]^{1/3}$, where $ g=
2t/(e^2/\epsilon\ell_0)$ and, from Figure \ref{ftfig2}, we estimate the
dimensionless constant $A \approx 30 $. (Unlike the results for $\lambda$ 
and $K$, the logarithmic
factor in the energy is apparent in our numerical results at small
Zeeman coupling strengths.) This numerical estimate of $A$ compares
reasonably well with previous estimates\cite{Sondhi,lili}
in which the result $ A \approx 
25 $ is obtained.  The consistency of these numerical results affords
confidence in their accuracy.

\section{Microscopic descriptions of single skyrmion states}
\label{sec:micro}

\subsection{Exact diagonalizations}
\label{sec:exact}

The microscopic physics of skyrmions can be approached directly by
numerically diagonalizing the finite-size Hamiltonian matrix for $N$
two-dimensional spin-1/2 electrons in $N_\phi$ lowest Landau level
orbitals.\cite{Sondhi,rezayi,he} The Hamiltonian commutes with the total
spin operator, $\hat S^2$, and, when the Zeeman term is neglected, also
with its projection, $\hat S_\xi$, on any direction $\xi$. When the Zeeman
coupling is present the direction $\xi$ must correspond with the direction
of the field. All the eigenstates of the system can be labeled by the
quantum numbers $S$ and $S_\xi$.  As mentioned in the Introduction, the
particular case $N=N_\phi$ corresponds to $\nu=1$ and, although there is
no rigorous proof, numerical evidence clearly indicates that the ground
state has the maximum possible total spin $S=N/2$ and, in the absence of
a Zeeman field, a degeneracy $2S+1$. As pointed out by Jain and
Wu\cite{jain}, this is one of the few cases where the first Hund's rule
applies in the fractional Hall regime. When the Zeeman field is in the
$\hat z$ direction, the ground state has $S_z=N/2$ and the electronic
system is completely spin-polarized in the direction of the field at
zero temperature.

\begin{figure}
\centerline{\epsfxsize=8cm \epsfbox{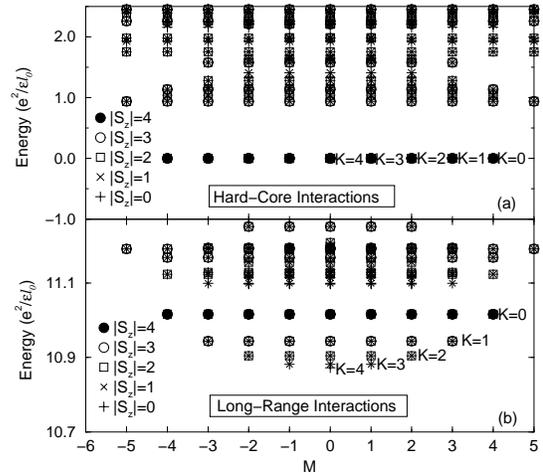}}
\caption{Energy spectrum obtained from an exact diagonalization of the
Hamiltonian in the spherical geometry with $N=8$ and $N_\phi=9$ for (a)
hard-core and (b) long-range (Coulomb) interactions (no Zeeman coupling
included). Different symbols have been used for different values of
$|S_z|$. For the case of the hard-core model a degenerate band
corresponding to the Landau level degeneracy occurs once for each member
of each $S = N/2 - K$ multiplet present at this system size.  For the
case of long-range interactions the degeneracy is removed and the energy
decreases as $K$($S$) increases(decreases). Rotational symmetry is
always present in the spherical geometry and the Landau level degeneracy
of the charged quasiparticles is preserved for each value of $K$.}
\label{jjpfig1}
\end{figure}

Since skyrmions are charged quasiparticles, they appear in exact
diagonalization calculations performed with $N=N_\phi \pm 1$.
(Particle-hole symmetry permits examination of one sign of charge only.)
In order to obtain the energy spectrum in the spherical
geometry\cite{Sondhi,rezayi,he} we need all interaction matrix elements
in the lowest shell of monopole harmonics\cite{monopole} which comprise
the lowest Landau level basis. These are given by 
\begin{eqnarray}
\langle m_1m_2|V(\vec\Omega-\vec\Omega')|m_3m_4\rangle = \frac{e^2}{\epsilon
\ell_0} (-1)^{m+m_1+m_2}&&\nonumber \\
\sum_{l=|m|}^{2N_S} \left(
 \begin{array}{ccc} N_S& l& N_S \\
 -m_1& -m& m_4 \end{array}\right)
\left( \begin{array}{ccc} N_S& l& N_S \\
-m_2& m& m_3 \end{array}\right) V_l^{N_S}&&,
\label{pseudo}
\end{eqnarray}
where $m=m_4-m_1=m_2-m_3$ and the $m_i$ are the azimuthal angular
momentum quantum numbers which label the single particle orbitals,
$V(\vec\Omega-\vec\Omega')$ is the interaction potential where
$\vec\Omega$ denotes the electronic spherical coordinate, and
$N_S=(N_\phi-1)/2$. The coefficients in this expansion are products of
Wigner's 3-j symbols and the $V_l^{N_S}$ parameters are the
interaction-dependent pseudopotentials introduced by Haldane. Since the
Hamiltonian also commutes with the total orbital angular momentum $\hat
L$ the diagonalization is carried out in subspaces of fixed $S_z$ and $M
\equiv L_z$. It is instructive to consider first the hard-core model for
the electron-electron interactions. In this case the pseudopotentials
$V_l^{N_S}$ are given by
\begin{equation}
V_l^{N_S}=\frac{(2N_S+1)^2(2l+1)}{\sqrt{N_S}}
\left( \begin{array}{ccc} N_S& l& N_S \\ -N_S& 0& N_S \end{array}\right)^2.
\end{equation}
For this model, as illustrated in Fig.\ \ref{jjpfig1}(a) and discussed
previously by Jain and Wu\cite{jain} and MacDonald, Fertig, and
Brey\cite{hcm}, multiplets of zero energy eigenstates occurs with all
possible values of the total spin quantum number $S$. In a quantum
description, the number of reversed spins in a skyrmion spin texture is
quantized and conjugate\cite{hcm} to its global orientation. The
zero-energy spin-multiplets with $S = N/2 - K$ have been shown\cite{hcm}
to be the quantum states which correspond to a classical skyrmion
texture with $K$ reversed spins. The results shown in Fig.\
\ref{jjpfig1}(b) are for the case of Coulomb interactions between the
spins, for which the pseudopotentials  are
\begin{equation}
V_l^{N_S}=\frac{(2N_S+1)^2}{\sqrt{N_S}}
\left( \begin{array}{ccc} N_S& l& N_S \\ -N_S& 0& N_S \end{array}\right)^2.
\end{equation}
In this case skyrmion states with more reversed spins have lower energy,
at least when the Zeeman energy is neglected. When the Zeeman energy is
included, the energy will in general be minimized at an intermediate
value of $K$. This is the result first obtained by Rezayi\cite{rezayi},
which provided the first evidence for the existence of charged
excitations with large spin quantum numbers in quantum Hall
ferromagnets. The appearance of degenerate states with different total
angular momenta reflects the extensive Landau level degeneracy of
single-skyrmion states with all values of $K$. 

\begin{figure}
\centerline{\epsfxsize=8cm \epsfbox{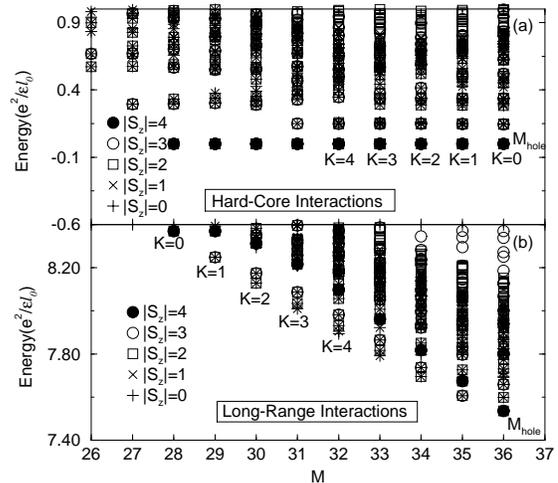}}
\caption{ Energy spectrum obtained from an exact diagonalization of the
Hamiltonian for a disk geometry with $N=8$ and $N_\phi=9$, for (a)
hard-core and (b) long-range (Coulomb) interactions (no Zeeman coupling
included). Different symbols have been used for different values of
$|S_z|$. As in Fig.\ \protect{\ref{jjpfig1}} for the case of the
hard-core model, a degenerate band at zero energy is observable where
all possible values of $K$ are present. Unlike in the spherical geometry
for the case of long-range interactions the translational symmetry is
broken due to strong edge effects.} 
\label{jjpfig2}
\end{figure}

We have also performed the same calculation ($N=8$, $N_\phi=9$) in the
disk geometry. Figure \ref{jjpfig2}(a) shows the low-energy spectrum for
the hard-core model where one can see the same degenerate band at zero
energy as in the spherical geometry case. For the zero-energy states of
the hard-core model the Landau level degeneracy is preserved. For the
Coulomb interaction, on the other hand, Landau level degeneracy is
broken by strong edge effects [see Fig.  \ref{jjpfig2}(b)] and the
non-local nature of the interaction. For a given value of $M$, the
minimum interaction energy state is always found at the maximum allowed
value of $K$.

\subsection{Hartree-Fock}
\label{sec:h-f}

A simpler and perhaps physically more transparent microscopic approach
for computing skyrmion properties is the unrestricted Hartree-Fock
approximation approach. This procedure finds the lowest energy Slater
determinant wavefunction which supports a spin texture of the form found
in skyrmion solutions of the NL$\sigma$ model. Such wavefunctions
generically have the form\cite{Herb} (for quasihole skyrmions)
\begin{equation}
|\Psi_{HF}\rangle = \prod_{m=0}^{\infty} (u_{m} c_{m\downarrow}^{\dagger} +
v_{m} c_{m + 1\uparrow}^{\dagger}) |0\rangle,
\label{slatdet}
\end{equation}
where $c_{m\sigma}^{\dag}$ creates a lowest Landau level electron in
angular momentum state $m$ with spin $\sigma$, and the $u_m,~v_m$'s are
variational parameters obeying the constraint $|u_m|^2+|v_m|^2=1$. The
precise way in which the energy of the state $|\Psi_{HF}\rangle$ may be
minimized with respect to the $u_m,~v_m$'s has been discussed
elsewhere\cite{Herb}.

In practical calculations, the Hartree-Fock approach bridges the gap
between skyrmion sizes which can be reached with exact diagonalization
approaches and systems sizes where the field theoretic approach becomes
accurate. Typically one is limited to system sizes with $N_\phi
\approx 10$ in
finite size diagonalization studies, since the matrix dimensions grow
rapidly with system size. On the other hand, Hartree-Fock calculations
are possible for spin textures involving up to several thousand
electrons. As in exact diagonalization studies one works with the exact
many-body Hamiltonian in Hartree-Fock studies so that the quantitative
errors introduced in the gradient expansion of the energy functional
used to generate the NL$\sigma$ model for the quantum Hall
ferromagnet\cite{kaneandlee,Sondhi,Moon} are not present. The
Hartree-Fock approach also captures some of the quantum mechanics of the
problem, though clearly not all. In particular, $|\Psi_{HF}\rangle$ has
$M-S_z$, the difference of the orbital and the $z$-component of the spin
angular momenta, as a good quantum number. However, $M$ and $S_z$ are
{\em not} separately quantized as they should be, nor is
$|\Psi_{HF}\rangle$ an eigenfunction of $\hat{S}^2$.  These failures
could easily have some quantitative importance for $K \sim 3$, the
quantum number of the lowest energy skyrmion in typical experiments.
From a field theoretic point of view, one may regard the Hartree-Fock
approximation as a mean field theory which retains the higher order
gradient terms absent in the simpler field theory studied in the
previous section. The resulting effective action would contain most
information, but, in principle, one should include quantum fluctuations
around this stationary phase approximation in order to properly account
for the quantization of orbital and spin angular momentum. As shall be
seen below, for the purposes of computing energies and sizes of the
skyrmions, the Hartree-Fock approach appears to be remarkably accurate,
suggesting that such quantum fluctuations are quantitatively
unimportant.  However, for probes sensitive to the quantum numbers of
the skyrmions -- tunneling spectroscopy for example -- wavefunctions
with the correct quantum numbers are needed to obtain qualitatively
correct results\cite{sky:tun}.

\subsection{Variational wavefunctions}
\label{sec:vw}

An alternate quantum description of single-skyrmion states can be
obtained by using variational many-body wavefunctions. These can be
refined by comparing with numerically exact solutions for small $K$ and
can provide a convenient, if somewhat indirect, route to calculate
properties of physical interest $N_\phi \to \infty$ limit.
The variational wavefunctions we
employ are motivated by the disk-geometry wavefunctions introduced by
MacDonald, Fertig and Brey\cite{hcm} which, in the thermodynamic limit,
become exact for hard-core-interaction single-skyrmion states. The
second quantized form of these wavefunctions clearly exhibits the
microscopic nature of the skyrmion:
\begin{eqnarray}
|\Psi^0_K\rangle=\frac{1}{\sqrt{C(K)}}\sum^{N_\phi}_{m_K>\cdots>m_1=1}
\frac{1}{\sqrt{m_1\cdots m_K}} && \nonumber \\ 
c^\dagger_{m_K-1\downarrow}c_{m_K\uparrow}\cdots
c^\dagger_{m_1-1\downarrow}c_{m_1\uparrow}c_{0\uparrow}|\Psi\rangle. &&
\label{skyr}
\end{eqnarray}
In the above expression $|\Psi\rangle$ denotes the ferromagnetic $\nu=1$ state
and $C(K)$ are the normalization constants
\begin{equation}
C(K)=\sum^{N_\phi}_{m_K>\cdots>m_1=1}\frac{1}{m_1\cdots m_K}.
\end{equation}
This wavefunction correspond to a skyrmion with quantum number $K$ located at
the origin of the plane with $M_{max}=M_{hole}-K$ ($M_{hole}$ corresponds to the
total angular momentum with a bare quasihole located at the origin [see Fig.\
\ref{jjpfig2}(a)]). All other single-skyrmion states with different orbital
angular momenta $M$ within a particular zero-energy $K$ band can be obtained by
lowering the center-of-mass angular momentum $n=M_{max}-M$ times:
\begin{equation}
|\Psi^n_K\rangle \propto
[\sum^{N_\phi}_{m=0}\sqrt{m+1}(c^+_{m\uparrow}c_{m+1\uparrow}
+c^+_{m\downarrow}c_{m+1\downarrow})]^n|\Psi^0_K\rangle.
\end{equation}
In a large system where edges may be neglected, all such states are degenerate.
Here we focus only on the states of highest orbital angular momentum, which are
centered at the origin.

It is interesting to note that these wavefunctions are identical to
hard-core model Hartree-Fock wavefunctions [Eq.\ (\ref{slatdet})]
projected onto a state of definite $S_z$\cite{wilczek}. (For the
hard-core model, the coefficients $u_m,v_m$ in Eq.\ (\ref{slatdet}) may
be determined analytically\cite{hcm}.) Thus, they presumably improve on
the Hartree-Fock wavefunction in that they include the quantum
fluctuations necessary to obtain the quantization of spin.  It should
also be noted that Eq.\ (\ref{skyr}) is not precisely an eigenstate of
$\hat{S}^2$, as may be verified by acting on it with the total spin
raising operator $S^+$. This operator should but fails to annihilate the
state for any finite size system. However, it is easily seen that
$S^+|\Psi^n_K\rangle \propto C(K)^{-1/2}$, which vanishes in the
thermodynamic limit. Thus, the relative weight of this wavefunction
outside its appropriate spin multiplet vanishes with increasing system
size.\cite{jacob}

\begin{figure}
\centerline{\epsfxsize=8cm \epsfbox{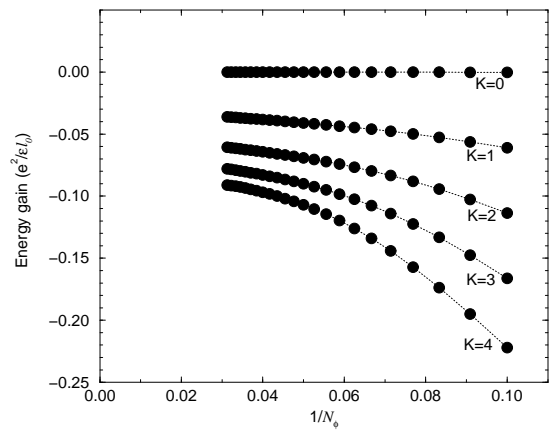}}
\caption{Energy gain of the skyrmion as a function of $1/N_\phi$ for
different values of $K$. These results were obtained using
unmodified hard-core model wavefunctions in the disk geometry (see
text) .}
\label{jjpfig3}
\end{figure}

\begin{figure}
\centerline{\epsfxsize=8cm \epsfbox{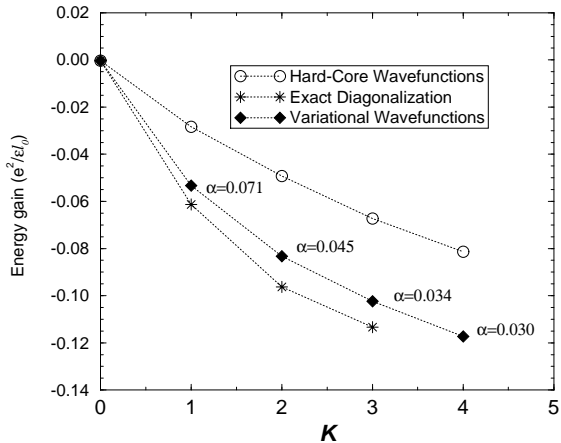}}
\caption{Energy gain of the skyrmion as a
function of $K$. The plot shows energies obtained using the
hard-core wavefunctions, the optimized variational wavefunctions,
and energies obtained from exact diagonalization in the spherical
geometry.  We were able to reliably extrapolate the exact
diagonalization results only for $K$ up to $3$. }
\label{jjpfig4}
\end{figure}

We use these variational wavefunctions to estimate the energies of single
skyrmion states for the case of Coulomb electron-electron interactions. Because
of numerical difficulties created by the long range of this interaction it is
convenient to introduce the neutral excitation energy defined as follows:
\begin{equation}
\epsilon(K)=\epsilon_+(0)+\epsilon_-(K)
\end{equation}
where $\epsilon_+(0)$ is the energy to make a $K=0$ antiskyrmion (bare
spin-reversed electron), 
\begin{equation}
\epsilon_+(0)=E(K=0,N=N_\phi+1)-E(N=N_\phi), 
\end{equation}
and $\epsilon_-(K)$ is the energy to make an skyrmion with $K$ reversed spins,
\begin{equation}
\epsilon_-(K)=E(K,N=N_\phi-1)-E(N=N_\phi).
\end{equation}
The quantity of physical relevance is the difference
$\epsilon(K)-\epsilon(0)$, the interaction energy gain resulting from a
texture with $K$ reversed spins.  The calculation of the neutral
excitation energies has been done for up to $N_\phi=32$ and the results
for this energy gain are shown in Fig.\ \ref{jjpfig3} as a function of
$1/N_\phi$. The edge effects, which bear primary responsibility for the
finite-size dependence seen in Fig.\ \ref{jjpfig2}, are minimized by
increasing the angular momentum of the minority-spin electron in the
$K=0$, $N=N_\phi+1$ state in step with $K$. This allows us to extract
reliable results from calculations at fairly small values of
$N_\phi$.\cite{caveat} In order to make an accurate extrapolation of
these results to the thermodynamic limit we use a Wynn's $\epsilon$
algorithm to generate Pad\'e approximants.\cite{wynn} The results of
this extrapolation are shown in Fig.\ \ref{jjpfig4}, where the same
method has been used to extrapolate the exact results obtained in
spherical geometry calculations (see Sec.\ \ref{sec:exact}) with
$N_\phi$ as large as $11$.

The substantial differences between exact diagonalization and
variational wavefunction results is indicative with deficiencies in
these wavefunctions. The source of the difficulty is that these
wavefunctions do not represent a finite size object as
$N_\phi\rightarrow\infty$. (This can be easily seen from the fact that
the normalization constants $C(K)$ diverge in the thermodynamic limit.)
{}From coefficients in the expression (\ref{skyr}) one can see that the
tail in the spin density distribution falls off like $1/r^2$, which
coincides with the long-distance behavior of the classical soliton
solution in the conventional $NL\sigma$ model for ferromagnets. However,
as stated in Sec.\ \ref{sec:fieldtheory}, the long-distance behavior
departs from this form when Coulomb and Zeeman terms are present. In the
classical solution the $1/r^2$ behavior is replaced by an exponential decay
at large distances. This observation motivates the following refinement
in our variational wavefunctions:
\begin{eqnarray}
|\Psi^0_K(\alpha)\rangle=\frac{1}{\sqrt{A(K)}}\sum^{N_\phi}_{m_K>\cdots>m_1=1}
\frac{e^{-\alpha(m_1+\cdots+m_K)}}
 {\sqrt{m_1\cdots m_K}}
&& \nonumber \\
c^\dagger_{m_K-1\downarrow}c_{m_K\uparrow} \cdots
c^\dagger_{m_1-1\downarrow}c_{m_1\uparrow} c_{0\uparrow} |\Psi\rangle &&,
\label{var1}
\end{eqnarray}
where the $A(K)$ are the new normalization constants
$$\sum^{N_\phi}_{m_K>\cdots>m_1=1}
\frac{e^{-2\alpha(m_1+\cdots+m_K)}}{m_1\cdots m_K}.$$ The parameter
$\alpha$ offers the possibility of optimizing the shape of the
skyrmion for each value of $K$.

It is, perhaps, surprising that the introduction of a Coulomb repulsion
leads to a more rapid fall off in the reversed spin density at large
distances from the skyrmion center. One way of understanding this
difference comes from examining the wavefunctions in Eqs.\ (\ref{skyr})
and (\ref{var1}). For $K = 1$ these are sums of products of a quasihole
in the $m=0$ orbital and a particle-hole pair with the minority-spin
electron precisely one unit of angular momentum lower than majority-spin
hole. Since such particle-hole pairs are precisely what are needed to
construct spin-waves at $\nu=1$\cite{kallin}, it is convenient to think
of the state as a spin wave with orbital angular momentum $m=-1$ moving
in the presence of a spin-polarized quasihole. For the case of hard-core
interactions, it is easy to see that the spin-wave disturbance and the
quasihole do not interact. The spin wave is therefore unbound and the
linear scale over which the spin disturbance is present is limited only
by the system size. For Coulomb interactions, the long-range nature of
the interaction leads to quite a different result. Because the
spin-minority electron is always closer to the origin than the
spin-majority hole, there is an effective {\em attractive} interaction
between the spin wave and the quasihole at the origin. The skyrmion thus
represents a state in which the spin-wave is bound to the quasihole so
that the spin disturbance is localized near the origin. Thus one should
expect the weight for large momentum particle-hole pairs to fall off
more quickly than for the case of the hard-core interaction.
Introduction of the parameter $\alpha$ provides the variational freedom
necessary for this to occur.

A deficiency of the states in Eq.\ (\ref{var1}) is that they are {\em
not} eigenfunctions of $\hat{S}^2$ (although they are, of course,
eigenstates of $S_z$). They can be modified to have correct quantum
numbers at a considerable cost in complexity as discussed in Ref.
\onlinecite{sky:tun}; interested readers are referred to that work for
details. The results we obtain using the present wavefunctions
extrapolating to the thermodynamic limit and the values of $\alpha$ that
optimize the energies are shown in Fig.\ \ref{jjpfig4}; the marked
improvement over the hard-core interaction wavefunctions (\ref{skyr}) is
apparent. As the value of $K$ is increased, the exponential factor
becomes irrelevant and one recovers the long-distance behavior expected
for classical solitons. This limit has been studied in Ref.\
\onlinecite{jacob} and Ref.\ \onlinecite{kamilla} and the result
obtained for the energy of the classical skyrmion\cite{Sondhi} is
reproduced by the hard-core wavefunctions.

\section{Comparison}
\label{sec:comparison}

\begin{figure}
\centerline{\epsfxsize=8cm \epsfbox{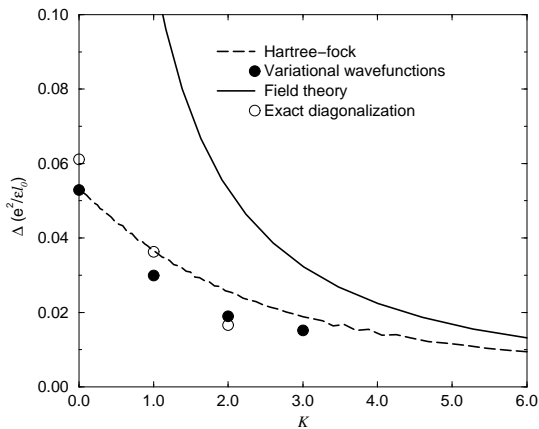}}
\caption{Skyrmion energy differences $\Delta (K)$
obtained from exact diagonalization (circles), variational wavefunctions
(black dots), the Hartree-Fock approximation (dashed line), and the
field theory (solid line). One can see the relative importance of
including higher-order gradient contributions (Hartree-Fock) and quantum
fluctuations (variational wavefunctions and exact diagonalizations).
$\Delta (K)$ is the Zeeman
spin-splitting value at which the number of reversed spins in the lowest
energy skyrmion changes from $K$ to $K - 1$.} \label{comfig1}
\end{figure}

This section is devoted to comparison of results obtained from the
classical field theory, Hartree-Fock approximation, and variational
wavefunction for Coulomb
interactions between electrons. In Fig.\ \ref{comfig1} we plot results
obtained for $\Delta (K) \equiv \epsilon(K) - \epsilon(K+1)$ calculated
with all three approaches. 

[For the classical field theory and the
Hartree-Fock approximation calculations, $K$ is not quantized and we
define $\Delta (K) \equiv d\epsilon /dK$.] $\Delta (K)$ is the value of
the Zeeman spin splitting ($ 2 t$ in the notation of this paper) energy
at which the single skyrmion states with $K$ and $K-1$ reversed spins
are equal in energy.  
For GaAs systems $\Delta (K)/ (e^2/\epsilon \ell)
\approx 0.006 \sqrt{ B [{\rm Tesla}]}$, depending somewhat on the
details of the semiconductor quantum well so that $K$ is typically $\sim
3$ for experiments performed at fields $\sim 10 {\rm Tesla}$. It is
clear from this plot that the corrections to the minimal field theory
captured by the Hartree-Fock approximation are sufficient to yield
accurate results. The use of leading gradient terms in the minimal field
theory is seriously in error for the small $K$ states of practical
interest. The additional quantum fluctuation effects included in the
variational wavefunction and exact diagonalization calculations are
never of overriding importance. 

\begin{figure}
\centerline{\epsfxsize=8cm \epsfbox{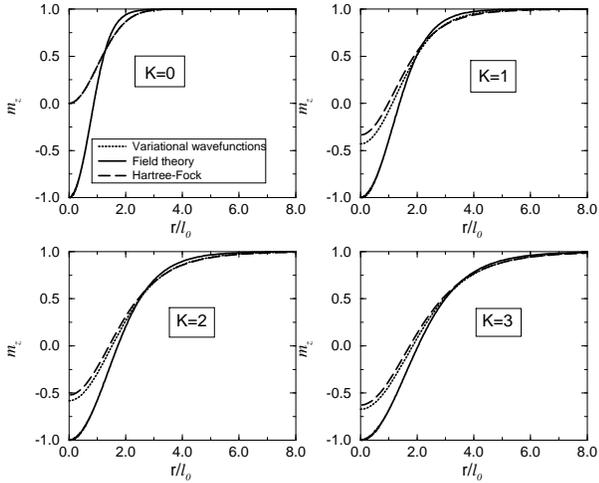}}
\caption{Radial distribution of $m_z$ in three approximations: classical
field theory (solid lines), variational wavefunctions (dotted lines), and
Hartree-Fock (dashed lines). The four smallest spin textures have been
considered: $K=0$(a), $K=1$(b), $K=2$(c), and $K=3$(d). For $K=0$ the
results of the variational wavefunctions and the Hartree-Fock
approximation lead to the same solution and, as expected, both schemes
converge to the classical field theory for increasing skyrmion size.}
\label{comfig2}
\end{figure}

Figure \ref{comfig2} compares the spin-textures by plotting $m_z(r)$ for
classical field theory, Hartree-Fock, and variational wavefunction
approaches at small integer values of $K$. For the classical field
theory and Hartree Fock approximations we plot the solution of Eq.\
(\ref{eq6}), which gives integer values for
the number of reversed spins $K$ as defined in Eq.\ (\ref{eqK}). In the
quantum calculations 
\begin{equation}
m_z(r)= 2\pi\ell_0^2 \; \sum_{m}\sum_{s=\pm 1} \; s \; n_m^s \; |\psi_m({\bf
r})|^2,
\label{eqM}
\end {equation}
where $n_{m\sigma}=\langle \Psi_K^0(\alpha)|c_{m\sigma}^\dagger
c_{m\sigma}|\Psi_K^0(\alpha) \rangle$ are single particle occupation
numbers, and $\psi_m({\bf r})$ are the angular momentum states of the
lowest Landau level in the symmetric gauge. Note that for the $K=0$
skyrmion state, the variational wavefunction and the Hartree-Fock
calculations are identical. In fact this state is a simple
spin-polarized quasihole state which is known to be given exactly by a
single Slater determinant. Apparently the fact that the Hartree-Fock
approximation becomes exact both for $K =0$ and $K \to \infty$ leads to
good accuracy at all values of $K$. As expected, the field theory
results approach those obtained with the variational wavefunctions and
Hartree-Fock approximations for large enough skyrmions while for small
skyrmions there are significant differences. The discrepancy is largest
for $K=0$ where the classical field theory calculation indicates that
the spin is reversed at the origin. In fact this property was one of the
boundary conditions used to solve the Euler differential equation of the
classical field theory. On the other hand, the quantum calculations show
that the z-component of the spin is zero at the origin; in fact the
total spin-density is zero at this point because the total charge
density is zero. These changes in the local properties are outside the
scope of the generalized NL$\sigma$ model energy functional.  These
errors are important for the smallest skyrmions but becomes less
significant with increasing skyrmion size.

On the other hand, the quantum exact diagonalization, variational
wavefunction and even Hartree-Fock calculations cannot be carried out
for very large skyrmions. The limit for the many-body calculations is $K
\sim 4$ which is adequate for comparing with existing experiments but
not for addressing asymptotic values of various quantities for the very
large skyrmions which would be obtained if the Zeeman coupling could be
adjusted experimentally to small values. Finite-size effects in these
quantum calculations are exacerbated by the long almost algebraic tails
of large $K$ skyrmion states. For example, Hartree-Fock calculations
with $N_\phi > 1000$ are necessary\cite{Herb} to obtain a accurate values of
$\Delta (K)$ near $K = 50$. Similarly it has not proven possible to
directly demonstrate the asymptotic behavior , $E \rightarrow
4\pi\rho_s$, as $t \rightarrow 0$ using Hartree-Fock calculations.
(However, the asymptotic value may be demonstrated by subtracting the
Hartree contribution to the Hartree-Fock energy and noting that this
should vanish as the skyrmion size diverges.) For the large skyrmion
states, which hopefully will be studied in future experiments, the
classical field theoretic approach is accurate and convenient. 

The discrepancy between the results of the classical field theory and
the microscopic calculations could be reduced by including quantum
fluctuations around the classical solution of the minimal field theory.
A uniform ferromagnet does not have quantum fluctuations around its
fully spin-aligned classical ground state and this is also the exact
quantum ground state. However, near a skyrmion center, the spin
direction changes with position and quantum fluctuations will occur.  As
the size of the skyrmion is reduced, these fluctuations will become more
important. It is clear that the sign of the effect will be to reduce the
spin polarization near the origin and bring the field theory results
into closer agreement with the microscopic results. Such calculations
have been done for the simple NL$\sigma$ model.\cite{Auerbach} However
computation of these fluctuations is numerically
difficult\cite{Abolfath}, even at the Gaussian level, in the present
model due to the non-locality of the Coulomb interaction. 

\section{Conclusions}
\label{sec:concl}

We have presented a study of single skyrmion states of quantum Hall
ferromagnets using and comparing a variety of techniques. We have
obtained the classical saddle point solution of the appropriate
non-linear sigma model modified to include the long-range Coulomb
interactions and Zeeman coupling. This minimal classical field theory
model includes the leading order gradients necessary to obtain stable
skyrmions in the presence of Zeeman coupling. We have developed and
evaluated variational wave functions describing states with the
appropriate quantum numbers. Finally, we have performed microscopic
numerical exact diagonalization calculations on finite size systems.
These results are compared with results obtained using a Hartree-Fock
approximation and reported in earlier work.

Our classical field theory results confirm that the skyrmion size scales
as the inverse cube root of the Zeeman energy as expected from the
competition between the Zeeman energy and the Coulomb energy and
predicted earlier\cite{Sondhi} by Sondhi {\em et al.}. We find that, for
skyrmion states with small numbers of reversed spins, the Hartree-Fock
results are in excellent agreement with many-body exact
diagonalization calculations and with calculations based on variational
wavefunctions motivated by exact zero-energy eigenstates of hard-core
model systems. Comparison of these calculations suggests that the
minimal classical field theory model is not quantitatively adequate for
the size of skyrmion which plays an important role in typical experiments in
GaAs quantum well systems. On the other hand quantum Hartree-Fock
calculations are shown to be in excellent agreement with many-body exact
diagonalization and variational wavefunction approaches.   Large
skyrmion systems may become experimentally available if the Zeeman
coupling strength can be tuned to smaller values, for example by
application of uniaxial stress to semiconductor host of the quantum
well, and disorder effects can be controlled. In this case quantum
calculations become impractical and the classical field theory approach,
possibly supplemented by field theory quantum fluctuation calculations,
may be necessary to describe experiments. 

\section{Acknowledgements}

The authors are grateful to D. Pfannkuche and L. Belkhir for useful
discussions.  Work at Indiana University and the University of Kentucky
was supported by the National Science Foundation under grants
DMR-9416906 and DMR 95-03814 respectively. MA acknowledges financial
support from The Center for Theoretical Physics and Mathematics, Tehran.
HAF acknowledges support from a Cottrell Scholar Award from the Research
Corporation.

%%%%%%%%%%%%%%%%%%%%%%%%%%%%%%%%%%%%%%%%%%%%%%%%%%%%%%%%%%%%%%%%%%%%%%%%%%%%%%%

\end{document}